\documentclass[Physsubmission, Phys]{SciPost}

\hypersetup{
    colorlinks,
    linkcolor={red!50!black},
    citecolor={blue!50!black},
    urlcolor={blue!80!black}
}

\usepackage[bitstream-charter]{mathdesign}

\DeclareSymbolFont{usualmathcal}{OMS}{cmsy}{m}{n}
\DeclareSymbolFontAlphabet{\mathcal}{usualmathcal}

\newcommand{\beqa}{\begin{eqnarray}}
\newcommand{\eeqa}{\end{eqnarray}}
\newcommand{\epm}{\begin{pmatrix}}
\newcommand{\bpm}{\end{pmatrix}}

\newcommand{\nn}{\nonumber}

\newcommand{\ii}{\text{i}}

\def\d{\mathrm{d}}

\begin{document}

\begin{center}{\Large \textbf{
Generalisation of affine Lie algebras on compact real manifolds\\
}}\end{center}

\begin{center}
Rutwig Campoamor-Stursberg\textsuperscript{1},
Marc de Montigny\textsuperscript{2} and
Michel Rausch de Traubenberg\textsuperscript{3$\star$}
\end{center}

\begin{center}
{\bf 1} Instituto de Matem\'atica Interdisciplinar and Dpto. Geometr\'\i a y Topolog\'\i a, UCM, E-28040 Madrid, Spain
\\
{\bf 2} Facult\'e Saint-Jean, University of Alberta, 8406 91 Street, Edmonton, Alberta T6B 0M9, Canada
\\
{\bf 3} Universit\'e de Strasbourg, CNRS, IPHC UMR7178, F-67037 Strasbourg Cedex, France
\\
* Michel.Rausch@iphc.cnrs.fr 
\end{center}

\begin{center}
\today
\end{center}


\definecolor{palegray}{gray}{0.95}
\begin{center}
\colorbox{palegray}{
  \begin{tabular}{rr}
  \begin{minipage}{0.1\textwidth}
    \includegraphics[width=20mm]{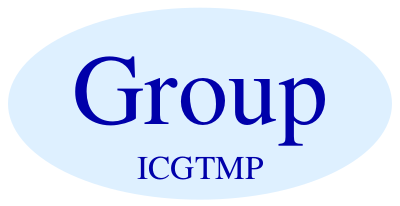}
  \end{minipage}
  &
  \begin{minipage}{0.85\textwidth}
    \begin{center}
    {\it 34th International Colloquium on Group Theoretical Methods in Physics}\\
    {\it Strasbourg, 18-22 July 2022} \\
    \doi{10.21468/SciPostPhysProc.?}\\
    \end{center}
  \end{minipage}
\end{tabular}
}
\end{center}

\section*{Abstract}
{\bf
We report on recent work concerning a new type of generalised Kac-Moody algebras based on the spaces of differentiable mappings from compact manifolds or homogeneous spaces onto compact Lie groups.

}

\vspace{10pt}
\noindent\rule{\textwidth}{1pt}
\tableofcontents\thispagestyle{fancy}
\noindent\rule{\textwidth}{1pt}
\vspace{10pt}

\section{Introduction}
\label{sec:intro}

Among the infinite-dimensional groups and algebras motivated by physical problems, the Virasoro, Kac-Moody, current and $W$-algebras and their representations are the most relevant representatives, and constitute a fundamental tool in several theories, such as Conformal Field Theory, gauge and string theories or SUGRA models (see \cite{GoL,west,ferrara} and references therein). It turns out that Kac-Moody algebras, as well as the associated Virasoro algebras, provide a natural framework for the unification of symmetry and locality properties \cite{Dol}. Basing on different physical assumptions, several generalisations of these algebraic structures have been proposed, usually from an analytic point of view, rather than on the axiomatic construction of these entities \cite{Mdo}. In this context, the quasisimple Lie algebras \cite{KT}, generalised Kac-Moody algebras based on geometrical properties of closed surfaces \cite{Frap} as well as several hierarchies of centrally extended algebras are worthy to be mentioned \cite{Borc,RS,RS2,jap}. 
 
 \medskip
 In most of these constructions, the one-dimensional sphere $\mathbb S^1$ plays a relevant role, a fact that suggests that, for other physical models involving more than one degree of freedom and related to some basis manifold, a similar procedure can be proposed, provided that the manifold is either compact or presents some peculiar properties than guarantee convergence of integrals. This situation was the starting point for the general procedure initiated in \cite{rmm}, where a systematic construction of generalised Kac-Moody algebras based on compact manifolds ${\cal M}$ 
related to either a Lie group or an appropriate homogeneous space was proposed. Under these assumptions, harmonic functions on the manifold can be described in terms of the representation theory of the corresponding Lie group, allowing us, in particular, to identify a complete set of Hermitean labelling operators. An important difference of this generalisation with respect to the well-known class of usual Kac-Moody algebras and other generalisations resides in the fact that our construction, based on the Fourier expansion on compact manifolds, does not imply in general the existence of simple roots, even if a root structure can  always be identified. 

  \medskip 
  Besides the interest of these generalised Kac-Moody algebras from the  mathematical point of view of,
 as this kind of algebras is naturally related to higher-dimensional compact manifolds,  the question of their relevance in theories involving higher dimensional space-times such as Kaluza-Klein theories, supergravity, {\it etc}  is  certainly of physical interest.
 
\section{The algorithmic construction of generalised Kac-Moody algebras\label{Sec4}}

The construction of generalised Kac-Moody algebras proposed in \cite{rmm} for the case of manifolds associated to either a compact Lie group $G_c$  or a coset space $G_c/H$ (via the exponential map, see \cite{rmm} for details) starts with a simple compact\footnote{This analysis can of course be extended to any simple (real or complex) Lie algebra. However, only in the case of compact Lie algebras, the representation theory has been analysed (see below).}
 Lie algebra $\mathfrak g$, a given basis $\{T_a, a=1,\cdots, \dim \mathfrak g\}$ with structure tensor 
\begin{equation}
\big[T_a,T_b\big] = \ii \;f_{ab}{}^c T_c \ , \nn
\end{equation}
and Killing form 
\begin{equation}\label{KIL}
\Big<T_a,T_b\Big>_0 = g_{ab} \equiv\  \text{Tr}\Big(\text{ad}(T_a) \text{ad}(T_b)\Big)  \ . 
\end{equation}

Denoting by $V$ the volume elements of the associated 
compact $n=(p+q)-$dimensional manifold ${\cal M}$ (with ${\cal M}\simeq G_c$ or ${\cal M}\simeq G_c/H$, we consider a local coordinate frame 
$ y^A = (\varphi^i, u^r)$ with $1\leq i\leq p$, $1\leq r\leq q$ such that the condition 
\beqa
\int _{\cal M} \d \mu({\cal M}) = \frac 1 V \int_{\cal M} \d ^p \varphi \; d^q u = 1 \ \nn
\eeqa
holds. On ${\cal M}$ we consider  the set of square integrable functions periodic in all $\varphi-$directions, but non-periodic in the $u-$directions. The space $L^2({\cal M})$ admits a  complete  orthonormal Hilbert basis 
\beqa\label{HiBa}
{\cal B} = \Big\{\rho_ I(\varphi,u)\ , \ \ I \in {\cal I} \Big\} \ ,
\eeqa
with respect to the Hermitean scalar product
on $L^2({\cal M})$, where ${\cal I}$ denotes a minimal (countable) set of labels required to identify the states
unambiguously \cite{C151}. In these conditions, we define a space of smooth mappings from ${\cal M}$ into $\mathfrak g$ as
\beqa
\mathfrak g({\cal M}) = \Big\{T_{a I} = T_a \rho_I(\varphi,u) \ , a = 1, \dots, \dim \mathfrak g\ , I \in  {\cal I} \Big\} \ . \nn
\eeqa
On this space, that inherits the structure of a Lie algebra, the Lie brackets are well defined and adopt the generic form 
\beqa
\label{eq:KM-def}
\big[T_{aI}, T_{bJ}\big] =\ii\; f_{ab}{}^c c_{IJ}{}^K T_{cK} \ ,
\eeqa
where the coefficients $ c_{IJ}{}^K $ are those of the Fourier expansion of products of elements in the basis ${\cal B}$. For the case where the manifold ${\cal M}$ is related to a compact Lie groups $G_c$, these can be associated to the Clebsch-Gordan coefficients of $G_c$. In particular, the Killing form in $\mathfrak g({\cal M})$ is given by
\beqa
\label{eq:killKM}
\Big<X,Y \Big>_1 = \int _{\cal M} \d \mu({\cal M}) \Big<X,Y \Big>_0  ,\quad X,Y \in \mathfrak g({\cal M}), 
\eeqa
from which the relations
\beqa 
\rho_I  (\varphi, u) = \eta_ {IJ} \overline{\rho}^ J (\varphi, u) \ , \quad 
\Big<T_{aI} ,T_{bJ}  \Big>_1 = g_{ab} \eta_{IJ} \  \nn
\eeqa
follow at once.  The first relation simply means that  $\overline{\rho}^ J \in L^2({\cal M})$, and thus extends in the basis ${\cal B}$ given by \eqref{HiBa}. 

\medskip
\noindent In a second step, the existence of central extensions for the preceding algebras is analyzed. Following a general approach based on cohomological methods, the central extension is obtained through the 2-cocycle  
\beqa
\label{eq:2Cocy-gamm}
\omega_C(X,Y) = \int_{{\cal M}}  \big<X, \partial_i Y\ \d \varphi^i + \partial_s Y\ \d u^s\big>_0 \wedge \gamma \ , 
\eeqa
with $\gamma$ being a closed $(n-1)$-current associated to the closed loop $C$. In this context, it should be taken into account that central extensions are associated to compact one-dimensional submanifolds of $\cal M$, {\em i.e.} curves, and that the procedure cannot be extrapolated to maps from higher-dimensional manifolds onto ${\cal M}$ \cite{ps}. Specifically, we consider 
\beqa
\gamma_{(A)} =    (-1)^A k_A  \d y^1 \wedge \cdots \wedge \d y^{A-1} \wedge  \d y^{A+ 1}\wedge \cdots \d y^n \ , \ A=1,\cdots ,n,\quad k_A \in \mathbb R. \nn
\eeqa
This leads to the identity 
\beqa
\label{eq:gam}
\omega_{(A)}(T_{aI}, T_{bJ})&=&  k_A g_{ab} \int _{{\cal M}} \d \mu({\cal M}) \; \rho_I(\varphi,u) \partial_A \rho_J(\varphi,u) = k_A g_{ab} d_{AIJ} \ ,
\eeqa
hence for the centrally  extended algebra $\mathfrak{g}({\cal M})$ we get the commutator 
\beqa
\label{eq:KM-ce}
\big[T_{aI},T_{bJ}\big] = \ii \; f_{ab}{}^c c_{IJ}{}^K T_{cK} +  g_{ab} \sum \limits_{A=1}^n k_A d_{AIJ} \ . 
\eeqa
It is not casual that this algebra has a deep similitude with the current algebra defined through 
\beqa
\label{eq:CA}
\big[T_a(y),T_{a'}(y')\big] = \ii \;f_{a a'}{}^b T_b(y) \delta^n(y-y') -\ii\; \sum \limits_{A=1}^n k_A \partial_A \delta^n(y-y') \ ,
\eeqa
and possessing Schwinger terms. Actually, centrally extended extensions of the generalised Kac-Moody algebras associated to the compact manifolds 
$\mathbb S^2$ and $\mathbb S^1 \times \mathbb S^1$ were determined in \cite{Bars} by means of current algebras, showing the validity of the procedure. 

\medskip
\noindent In a third step, derivations $\partial_A$ of the generalised Kac-Moody algebra $\mathfrak{g}({\cal M})$ are considered. This is a technically delicate step, as the variables $\varphi$ are periodic, whereas the variables $u$ do not exhibit periodicity properties. In other words, the operators $d_j=-\ii\partial_{\varphi^j}$ are (commuting) Hermitean, while the operators $d_s= -\ii \partial_{u^s}$ are not Hermitean. In order to obtain a complete set of commuting Hermitean operators, we use the identification of the manifold ${\cal M}$ with a compact Lie group (coset space). To this extent, an embedding $\mathfrak g_c \subseteq \mathfrak g_m$ of $\mathfrak g_c$ into a higher-rank Lie algebra $\mathfrak g_m$ is used, with $\mathfrak g_c$ the Lie algebra of $G_c$, and such that the basis functions of (\ref{HiBa}) belong to an irreducible  unitary representation of $\mathfrak{g}_m$. 
The generators of the latter can be realised as Hermitean differential operators acting naturally on the manifold; in particular, the elements $h_1,\cdots, h_k$ of the Cartan subalgebra of $\mathfrak g_m$ (where $k$ is the rank of $\mathfrak g_m$), are realised as the Hermitean operators 
\beqa
h_j= -\ii f_j^A(y) \partial_A, \quad 1\leq j\leq k. \nn
\eeqa
A particularity of these operators is that the boundary term associated to all $u-$directions vanishes. Among the operators $\left\{  d_1, \cdots, d_p, h_1,\cdots, h_k\right\}$ we determine a maximal set of commuting operators 
\beqa
D_j=-\ii f_j^A(y) \partial_A,\quad \ j=1,\cdots, r \nn
\eeqa
that satisfy the constraints 
\beqa
\label{eq:herm}
\partial_A f_j^A (y) =0 \ \  \text{and}\ \  f_j^r| =0\ , \ j=1,\cdots, r\ ,
\eeqa
related to Hermiticity. In these expressions,  $f_j^r|$ represent the boundary terms associated to all  $u-$directions that must vanish. Note further that when ${\cal M} = \mathbb T^n$, as all directions are periodic, we have $r=n$, but for a generic $n-$dimensional manifold ${\cal M}$ we have $r<n$. It can be easily shown that the $D_i$ act diagonally on the functions $\rho_I$, leading to an eigenvalue problem  
\beqa\label{EIW}
D_j(\rho_I(y))= I(j) \rho_I(y) \ , \nn
\eeqa
with $I(j)$ the corresponding eigenvalue.

\medskip
\noindent The Hermitean operators $D_j$ and central extensions of the generalised Kac-Moody algebra are deeply related  through 
the closed $(n-1)-$forms ($j=1,\cdots, r$)
 \beqa
\label{eq:form}
\gamma_j =  k_ j\sum \limits_{A=1}^n (-1)^A f_j^A(y)\; \d y^1 \wedge \cdots  \wedge \d y^{A-1} \wedge \d y^{A+1} \wedge \cdots \wedge \d y^n \ ,\ 
 j=1,\dots,
r,\quad k_j\in \mathbb R, \
\eeqa
with corresponding 2-cocycles given by   (see equation \eqref{eq:gam})
\beqa
\label{eq:cc}
\omega_k(T_{aI},T_{bJ}) =  k_k  J(k) g_{ab} \eta_{IJ}\ . 
\eeqa

\medskip
\noindent
Summarising, the generalised Kac-Moody algebra $\widehat{\mathfrak g}({\cal M})$  associated to the compact Lie algebra $\mathfrak g$ and the compact manifold ${\cal M}$ is determined by the following data

\begin{enumerate}
\item  Generators $T_{aI}$ belonging to $\mathfrak{g}({\cal M})$;
\item  Commuting Hermitean operators $D_1,\cdots, D_r$;
\item Central charges $k_1,\cdots, k_r$ associated to the Hermitean operators.
\end{enumerate}
If $I(j)$ denotes the eigenvalue of $D_j$ (see (\ref{EIW})), the non-vanishing brackets of the generalised Kac-Moody algebra associated to ${\cal M}$  are 
\beqa
\label{eq:KM-gen}
\big[T_{aI},T_{bJ}\big] &=&\ii\;  f_{ab}{}^c c_{IJ}{}^K T_{cK} +    g_{ab}  \eta_{IJ}  \sum \limits_{j=1}^r k_j I(j)
 \ , \nn\\
\big[D_j, T_{aI}\big] &=&   I(j)  T_{aI}\ ,
\eeqa
where $I(j)$ is the eigenvalue of $D_j$. Recall again that the central charges and the Hermitian operators are both   associated to the closed $(n-1)-$form given by \eqref{eq:form}.

\medskip
\noindent As shown in \cite{rmm}, the choice of $G_c = U(1) ^n$ leads to a generalised Kac-Moody algebra that structurally coincides with the generalised algebras based on the torus $\mathbb T^n$ studied and analyzed in \cite{KT}, and that actually correspond to specific cases of the wide class of so-called `quasi-simple Lie algebras'.   

\section{Identification of roots in $\widehat{\mathfrak g}({\cal M})$}  

\noindent The fourth step is devoted to the identification of a root structure based on equation \eqref{eq:KM-gen} associated to generalised Kac-Moody algebras. Supposed that the initial simple Lie algebra $\mathfrak g$ has rank $\ell$ and let $\Sigma$ denotes the root system with respect to the Cartan subalgebra 
$H^i, i =1,\cdots, \ell$, we consider the root operators $E_\alpha, \alpha \in \Sigma$ in the Cartan-Weyl basis. Defining 
\beqa
\label{eq:hatg}
\hat{ \mathfrak g}({\cal M})= \text{Span}\Big\{ T_{aI}, D_j, k_j, a=1,\cdots, \dim \mathfrak g, I \in {\cal I}, j=1,\cdots, r \Big\} \ ,
\eeqa
the Cartan subalgebra  of the latter is spanned by $H^i$, $D_j$ and $k_j$ ($i=1,\dots, \ell,\; j = 1,\cdots, r$). Taking  the Cartan-Weyl basis $H^i_I$, $E_{\alpha I}$ and the Killing form as defined in (\ref{eq:killKM}), application of the procedure described in
\cite{go} shows that the Killing form of $\hat{ \mathfrak g}({\cal M})$ satisfies 
\begin{equation}\label{eq:CSA}
\begin{split}
\Big<T_{aI},T_{bJ}\Big>& =\eta_{IJ} g_{ab},\quad 
\Big<D_j,T_{aI}\Big>=\Big<k_ j ,T_{a I }\Big> \ =\ 0\ , \\
\Big<k_i ,k_j\Big> &=\Big<D_i ,D_j\Big>\ =\ 0,\quad 
\Big<D_i ,k_j\Big>=\delta_{j}^{i}.
\end{split}
\end{equation}
From this we get the (infinite-dimensional) root spaces (where ${\bf n } = (n_1,\cdots,n_r)$)
\begin{equation}\label{eq:root-s}
\begin{split}
\mathfrak g_{(\alpha, {\bf n })} &= \Big\{E_{\alpha I } \ \text{with} \ I(1) = n_1 ,\cdots , I(r)=n_r \Big\} \ ,\ \ \alpha \in \Sigma\ , \ \ 
n_1,\cdots, n_r \in \mathbb Z, \\ 
\mathfrak g_{(0, {\bf n } )} &= \Big\{H^i_{I}\  \text{with} \ I(1) = n_1 ,\cdots , I(r)=n_r \Big\} \ , \ \ n_1,\cdots, n_r \in \mathbb Z \ , 
\end{split}
\end{equation}
with commutation relations 
\beqa
\big[\mathfrak g_{(0,{\bf n})}, \mathfrak g_{(\alpha,{\bf m})}\big]&\subset& \mathfrak  g_{(\alpha,{\bf m+n})}, \nn\\
\big[\mathfrak g_{(\alpha,{\bf m})}, \mathfrak g_{(\beta,{\bf n})}\big]&\subset& \mathfrak  g_{(\alpha+\beta,{\bf m+n})}, \ \ \alpha+ \beta \in \Sigma\nn
\eeqa
An important difference with respect to the usual Kac-Moody algebras is that, in this case, the commutator between elements depends also on the representation theory of $G_c$, specifically in connection with the Clebsch-Gordan coefficients $c_{IJ}{}^K$. This shows that the construction goes beyond the traditional root theory, as it also involves the so-called labelling problem for embedded algebras \cite{C151}. 

\medskip
Explicit construction of these generalised structures where obtained in \cite{rmm} for the case of manifolds isomorphic to the spheres $\mathbb{S}^n$, specifically for the values $n=2$ with $SU(2)/U(1)$, $n=3$ for $SU(2)$ and $SO(4)/SO(3)$, $n=5$ for $SU(3)/SU(2)$ and $n=6$ for $G_2/SU(3)$. 

\medskip
Concerning the representation theory of generalised Kac-Moody algebras, the case of the $n$-dimensional torus $\mathbb{T}^n$ has been inspected in some detail in \cite{rmm}, corresponding to the Lie group $U(1)^n$. An extrapolation to other more complicated manifolds is a delicate task, the technical difficulties of which have not yet been solved satisfactorily. However, for the two-dimensional case and the manifolds $\mathbb{T}^2$ and $\mathbb{S}^2$, an alternative ansatz has been proposed in \cite{Fermion} and \cite{Boson}, based on the observation that the Kac-Moody and the corresponding Virasoro algebras associated to these manifolds can be constructed naturally from the usual Kac-Moody and Virasoro algebras.
More specifically, in this case  we have assumed that the  Laurent modes
        of the usual Kac-Moody and Virasoro  algebras can themselves be (Fourier) developed  in an adapted manner on the two-sphere and the two-torus, respectively \cite{Fermion}. This assumption enabled us to reproduce the generalised Kac-Moody algebras associated to  $\mathbb S^2$ or $\mathbb T^2$,
      in a semi-direct product with  a subalgebra of vector fields of the two-torus and the two-sphere
        \beqa
        \label{eq:V-KM}
\text{Vir} ({\cal M}) \ltimes \widehat{g}({\cal M}) \ \ \ \text{with} \ \ \ {\cal M} = \mathbb S^1 \ \ \text{or} \ \ \mathbb T^2 \ .
        \eeqa
 The algebras  $\text{Vir} ({\cal M})$ can been seen as extensions of the Virasoro algebras in these cases. The interesting observation of this construction is that it leads naturally to central extensions.
The fermions \cite{Fermion} and boson \cite{Boson} realisations subsequently obtained lead automatically to a Fock space construction and thus to a unitary representation bounded from below. In the case of the bosonic construction, we have introduced vertex operators along the lines of the vertex operator in string theory \cite{fk, seg}.

\section{Concluding remarks and future prospects}

We have reported on recent work concerning on the construction of generalised Kac-Moody algebras for the class of compact Lie groups and certain coset spaces determined by a closed subgroup, and the analysis of some of its main features that may be of interest in physical applications, such as the existence of a root system and central extensions. 
The procedure can formally be developed for any compact manifold or homogeneous space of the specified type, with the main difficulties being of computational nature. Whether this class of extensions fits naturally in the description of physical phenomenology, is a problem that has still to be explored in more detail. 

\medskip
The next natural step, besides specific applications, consists in proposing an analogous construction for the case where the basis manifold $\cal M$ is no more compact. Some results in this direction actually exist, such as the work \cite{RS}, but a general approach has not been formulated yet. Among the obstructions observed in this general frame, the acute divergence problems that arise in the integration theory of non-compact manifolds, as well as the technical difficulties emerging in the cohomological formulation of central extensions (see equation \eqref{eq:2Cocy-gamm}), are the most relevant. Inspection of several examples suggest that additional techniques have to be considered to cover this case appropriately, in order to obtain a description the validity of which is not restricted to very particular manifolds. A successful approach in this sense could possibly be of interest in the context of $M-$theory or supergravity models, studying whether there is a connection between the central extensions of the generalised Kac-Moody algebra and super-membrane solutions in extended SUGRA. 

The fermionic \cite{Fermion} and bosonic \cite{Boson} construction obtained in the case of $\mathbb S^2$ and $\mathbb T^2$ can be easily extended to the $n-$tori $\mathbb T^n$. This extension leads to a hierarchy of algebras (with the notations
  of \eqref{eq:V-KM})
  \beqa
  \text{Vir}(\mathbb T^n) \ltimes \widehat{g}(\mathbb T^n)\subset \text{Vir}(\mathbb T^{n-1}) \ltimes \widehat{g}(\mathbb T^{n-1})
  \subset \cdots \subset \text{Vir} \ltimes\widehat{g}, 
  \eeqa
  where at the last stage we have the usual Virasoro and Kac-Moody algebras. This series of embeddings could play a role in toroidal compactifications,  a very important notion in  higher dimensional supergravity.

\section*{Acknowledgements}

\paragraph{Funding information}
RCS  acknowledges financial support by the research
grants MTM2016-79422-P (AEI/FEDER, EU) and PID2019-106802GB-I00/AEI/10.13039/501100011033 (AEI/ FEDER, UE).  MdeM is grateful to the Natural Sciences and Engineering Research Council (NSERC) of Canada for partial financial support (grant number RGPIN-2016-04309).

\bibliography{ref.bib}

\nolinenumbers

\end{document}